# Photonic integrated circuits in the continuum

Zejie Yu[1], Xiang Xi[1], Jingwen Ma[1], Hon Ki Tsang[1], Chang-Ling Zou[2] and Xiankai Sun[1,*]

[1]Department of Electronic Engineering, The Chinese University of Hong Kong, Shatin, New Territories, Hong Kong.

[2]Key Laboratory of Quantum Information, University of Science and Technology of China, Hefei, Anhui 230026, China.

[*]e-mail: xksun@cuhk.edu.hk

**Waves that are perfectly confined in the continuous spectrum of radiating waves without interaction with them are known as bound states in the continuum (BICs). Despite recent discoveries of BICs in nanophotonics, full routing and control of BICs are yet to be explored. Here, we experimentally demonstrate BICs in a fundamentally new photonic architecture by patterning a low-refractive-index material on a high-refractive-index substrate, where dissipation to the substrate continuum is eliminated by engineering the geometric parameters. Pivotal BIC-based photonic components are demonstrated, including waveguides, microcavities, directional couplers, and modulators. Therefore, this work presents the critical step of photonic integrated circuits in the continuum, and enables the exploration of new single-crystal materials on an integrated photonic platform without the fabrication challenges of patterning the single-crystal materials. The demonstrated lithium niobate platform will facilitate development of functional photonic integrated circuits for optical communications, nonlinear optics at the single photon level as well as scalable photonic quantum information processors.**

Solved from the Schrödinger equation for a particle in a finite potential well, the wavefunctions with the corresponding eigenenergy inside the potential well can be square integrable, and they are known as the bound states of the quantum system. Some other states, with their eigenenergy above the potential well and the wavefunction spreading to infinity, are known as continuous states. If the eigenenergy of a bound state is outside the band of the continuous states, e.g., the potential of the well is below that of the environment, the wavefunction is localized without energy dissipation. Using the similarity between the Schrödinger equation and the Helmholtz equation[1,2], the refractive index $n$ of a material determines the potential for photons as $-n^2 k_0^2$, where $k_0$ is the light's wave number in the vacuum. Therefore, a structure that supports bound photonic states should be made of a material that has a refractive index higher than the environment. The photonic architectures following this rule have laid the foundations for photonics and optoelectronics, such as optical fibers and photonic integrated circuits (PICs) that have been supporting telecommunications, information processing, and quantum technologies.

In contrast to this conventional wisdom, bound states in the continuum (BICs), which were first proposed by von Neumann and Wigner in 1929[3], refer to a type of eigenstates whose wavefunctions are square integrable with the corresponding eigenenergy above the potential well. Such counterintuitive phenomena are of fundamental importance in quantum mechanics[4] and have been discovered in acoustics[5-9], electronics[10-15], and photonics[16-29]. Although the BIC mechanism promises the confining and routing of photons without using a high-refractive-index material surrounded by low-refractive-index environment, the existing photonic BICs are all realized in the conventional architectures. Recently, the photonic BICs that exist in low-refractive-index PICs on a high-refractive-index substrate have been theoretically proposed [19]. Such a new configuration of photonic BIC breaks through the fundamental limitation of photonic architectures, allowing the exploration of high-refractive-index dielectric materials on an integrated platform without the need of patterning micro- or nanostructures in these materials.

**Results**

Here, we experimentally demonstrate a fundamentally new architecture of PIC, where the photons in a propagating bound mode are laterally confined, routed, and manipulated in the continuum. The propagation loss of the bound photonic mode in the high-refractive-index substrate is minimized by engineering the geometry of the microstructures made in the low-refractive-index material. Various types of photonic devices for this new type of PIC, which are essential for constructing on-chip functional PIC networks, are experimentally demonstrated. Integrated microcavities also exhibit the BIC effect and achieve ultrahigh quality ($Q$) factors, which is the first demonstration of cavities with three-dimensional (3D) confinement in the continuum to the best of our knowledge. This new PIC architecture by the BIC mechanism is simple and generic, allowing PICs with flexible choice of materials, such as lithium niobate ($LiNbO_3$), yttrium iron garnet (YIG), barium borate (BBO), yttrium aluminum garnet (YAG), and diamond, which are single-crystal dielectrics with excellent optical properties but challenging in nanofabrication.

Figure 1a is a conceptual illustration of the PIC in the continuum, where the PIC is constructed by patterning a low-refractive-index material atop instead of patterning the high-refractive-index single-crystal substrate (e.g., $LiNbO_3$-on-insulator wafer), and photons can be concentrated predominantly in the single crystal (e.g., $LiNbO_3$). The mechanism of lateral confinement of light in a channel waveguide is a conventional higher-effective-index channel in the $LiNbO_3$ as illustrated in the effective refractive index distributions in Fig. 1b (also see Supplementary Information). The low-refractive-index material is patterned in the lateral direction ($y$ direction), renders a high-effective-index channel for the transverse confinement of the transverse-electric (TE) and transverse-magnetic (TM) polarizations. Careful engineering is required to ensure that the optical power in the fundamental TM mode (i.e., the bound mode as shown in Fig. 1e), whose eigenfrequency lies in the continuum of the TE modes (Fig. 1d), does not dissipate to the TE modes. Due to the broken translational symmetry in the $z$ direction, the TM bound mode can interact with the TE continuous modes, resulting in energy dissipation of the former. Figure 1c shows a typical dispersion diagram of a straight waveguide for both the TM bound mode (blue surface) and the TE continuous modes (red surface). The intersecting line of the two surfaces indicates the phase matching between the TE continuous modes and the TM bound mode, which means that the TM bound mode embedded in the TE continuous spectrum can always match the dissipative waves in the continuum, leading to high loss for the TM bound mode.

In this paper we harness the BIC mechanism to avoid the dissipation of photons in the TM bound mode to the TE continuum through the destructive interference between different coupling channels. The attenuation length of the TM bound mode in a straight waveguide can be expressed as $L \propto w^2/\sin^2(k_y w/2)$, with $k_y$ the $y$ component of the wave vector of the TE continuous mode which matches that of the TM bound mode[19]. When $k_y w$ is a multiple of $2\pi$, $L$ approaches infinity and the BIC is obtained correspondingly. In Fig. 1c, the green dot and triangle represent respectively the non-BIC and BIC conditions, where their modal profiles are plotted in Figs. 1f and 1g, respectively. For a non-BIC state (Fig. 1f), the electric field extends in the substrate, resulting in dissipation of the propagating photons. In contrast, a BIC state (Fig. 1g) exhibits only localized electric field in the substrate, resulting in lossless propagation of the photons in the continuum.

To demonstrate the key features of BIC in our PIC architecture, we first consider the elementary photonic structures. All the numerical and experimental results in this paper and the Supplementary Information are based on a LiNbO$_3$-on-insulator wafer structure as illustrated in Fig. 1a, where the thicknesses of the LiNbO$_3$ layer, the buried silicon oxide underneath, and the organic polymer atop are 300 nm, 2 μm, and 500 nm, respectively, unless otherwise noted. Figure 2b shows the loss rate of the TM bound mode propagating in a straight waveguide (Fig. 2a) as a function of the waveguide width $w$ and the wavelength $\lambda$. The loss reduces dramatically at certain geometric parameters where the TM bound mode is totally decoupled from the TE continuous modes, becoming the desired BIC. As shown in Fig. 2c, the tolerance of waveguide width to maintain the loss rate below 1 dB/cm is as large as 300 nm, which can confidently be achieved with state-of-the-art nanofabrication techniques. Waveguides with a larger width also allow small propagation loss of the TM bound mode due to weak coupling to the TE continuous modes, albeit at the expense of looser confinement of photons. In addition, the TM bound mode can maintain ultralow loss in a large wavelength range as shown in the inset of Fig. 2c. This is attributed to small structural dispersion as the light is mainly confined to the LiNbO$_3$ thin film that extends in the $y$ direction.

To route photons on a chip, bent waveguides (Fig. 2d) are indispensable for joining various elements of PICs and for constructing microcavities. Similar to their straight counterparts, the interaction between the bound and continuous modes originates from the waveguide edges. Dissipation of the TM bound mode to the TE continuum in the high-refractive-index slab can be estimated from $|J_q(nk_0R) - \xi J_q(nk_0(R - w_b))|$, where $J_q$ is the $q$th Bessel function with $q$ the azimuthal mode number [19]. Here, the coupling to the continuum not only depends on the geometric parameters (waveguide width $w_b$ and bend radius $R_b$), but also is related to $\xi$, the ratio of electric field intensity at the two edges of the waveguide. Figure 2e shows the bending loss at 1.55 μm as a function of $w_b$ and $R_b$. The BIC can be obtained for certain combinations of parameters (dark regions), where near-zero bending loss can be achieved with large fabrication tolerance.

Microcavities of high optical $Q$ factors play an important role in many integrated photonic applications. For microring cavities, the bending loss of the propagating TM bound mode is the same as that for the bent waveguides. For microdisk cavities, there is only one edge where the dissipation occurs, so the bending loss of the propagating TM bound mode can be estimated by $|J_q(nk_0R)|$. Figure 2f shows the intrinsic $Q$ factor of microdisk cavities as a function of the disk radius. The $Q$ factor approaches infinity when the BIC condition is satisfied. With the capability of achieving ultrahigh $Q$ factors, the microcavities in the continuum also provide a scheme of controlling the $Q$ factors without affecting other on-chip devices fabricated in the same run. These cavities can maintain an ultrahigh $Q$ factor ($>10^5$) with large tolerance of geometric parameters and wavelength near the conditions for BICs (see Supplementary Information). It should be noted that although the loss to the substrate continuum can be eliminated by using the principle of BIC, the typical loss to the free space cannot be avoided. This is consistent with the no-go theorem for the BIC in compact photonic structures[30]. Increasing exponentially with the bend radius, the bending-loss-limited $Q$ factor can be regarded as infinity when the bend radius is sufficiently large. Therefore, when the BIC is obtained in practical experiments, the dissipation of photons will be limited only by the material absorption and fabrication imperfection, rather than the radiation loss to the substrate.

To experimentally demonstrate the BICs in such a PIC architecture, we fabricated and characterized the basic elements of PICs on a LiNbO$_3$-on-insulator platform. A 500-nm-thick e-beam resist (ZEP520A) was adopted as the upper low-refractive-index material such that only one step of e-beam lithography was sufficient and no further etching process was required for the fabrication. The photonic devices were characterized by coupling light from an optical fiber into and out of the PICs via grating couplers. The grating couplers were designed to be polarization sensitive so that they also served as mode selectors[31], enabling highly efficient excitation of the fundamental TM bound modes in the PICs. The propagation loss of straight waveguides was extracted from the transmission of waveguides with different lengths, as shown in Fig. 3a. The measured propagation loss (Fig. 3b) shows oscillation as a function of the waveguide width $w$ and reduces to near zero (two orders of magnitude lower than the maxima) for certain parameters, which agrees well with the theoretical prediction for the BICs (Fig. 2b). The bent waveguides were characterized by measuring the transmission from the devices as shown in Fig. 3c, from which the bending loss was extracted (see Supplementary Information). The bending loss as a function of the bend radius $R_b$ and waveguide width $w_b$ is plotted in Figs. 3d and 3e, respectively, which agrees with the theoretical prediction for the BICs (Fig. 2e). To demonstrate 3D confinement of photons based on BIC, we also fabricated microdisk and microring cavities. The modal properties were characterized by measuring the transmission spectra from a waveguide evanescently coupled with the cavity. The dependence of the intrinsic $Q$ factors of cavity resonance around 1,550 nm on disk radius $R_d$ (Fig. 3g) shows excellent agreement between the simulated and experimental results, which confirms the achievement of BIC in photonic microcavities. The $Q$ factor saturates at a finite value, which is attributed to the material absorption (see Supplementary Information). Meanwhile, the measured transmission spectrum of an undercoupled microring cavity fabricated on a 400-nm $LiNbO_3$-on-insulator wafer (Fig. 3h) indicates that the demonstrated microcavities can support high-$Q$ resonances with high extinction over a broad bandwidth. The zoomed-in spectra in Fig. 3i indicate the loaded and intrinsic $Q$ factors are $2.2 \times 10^5$ and $5.8 \times 10^5$, respectively.

In contrast to the photonic BICs in 1D or 2D configurations, the demonstrated BIC in our new architecture allows for construction of fully functional PICs. In addition to waveguides and microcavities, the multiport components including directional couplers (Fig. 4a), Mach–Zehnder interferometers (MZIs, Fig. 4c), and electro-optic modulators (Fig. 4e) are the kernel components of a scalable PIC. The transmission spectra of a fabricated directional coupler (Fig. 4b) indicate that light can efficiently be delivered from one port to another in the wavelength range of 1,450–1,600 nm. Such a broad operating bandwidth is an advantage of the BIC-based PIC architecture, which is attributed to the small structural dispersion. Figure 4d shows the measured transmission spectrum of a MZI (Fig. 4c), which is constructed from two 3-dB power splitters. The length difference between the two arms of the MZI leads to the beating pattern in the spectrum, where the large extinction ratio indicates that the 3-dB power splitters in the MZI can divide optical power evenly. $LiNbO_3$ is well known for its large electro-optic coefficients, which allow for active tuning and reconfiguration of PICs. As a proof of concept, we fabricated electrodes together with the MZI to make an electro-optic modulator as shown in Fig. 4e, where the inset is a scanning electron microscope (SEM) image zoomed in at the electrodes. When a voltage is applied to the electrodes, the refractive index of $LiNbO_3$ changes, leading to a variation of the effective refractive index of the BIC. Consequently, we can tune the output power of the MZI due to the varied phase difference of the two arms. The minimal voltage for completely switching the output power between on and off is referred to as the half-wave voltage ($V_\pi$), which is ~16 V for a 5-mm-long MZI modulator with an extinction ratio of 15 dB (Fig. 4f). Figure 4g shows the temporal response of the MZI electro-optic modulator, indicating the modulation speed can be higher than 100 Mbps (see Supplementary Information). Such devices can be used for manipulating the frequency properties of photons on a chip, and the electro-optic interaction can be enhanced further in high-$Q$ microcavities, which is promising for achieving the microwave-to-optical frequency conversion[32].

**Discussion**

The demonstrated PICs in the continuum break the refractive index limitation of materials, and thus enable a new class of integrated photonic devices where photons are confined to low-refractive-index microstructures on high-refractive-index substrates. Due to the generic mechanism of BICs, the hybrid photonic architecture consisting of polymer on single-crystal thin film[19] presented here can be extended directly to other functional materials, which promises a solution to the key challenges encountered in traditional integrated photonics. On one hand, polymers are compatible with dye molecules as well as quantum dots, allowing the realization of on-chip optical amplifiers and low-threshold lasers at various wavelengths. On the other hand, single-crystal materials can provide excellent linear and nonlinear optical properties. The lithium niobate platform demonstrated in this work will facilitate development of functional photonic integrated circuits for optical communications, nonlinear optics at the single photon level as well as scalable photonic quantum information processors. For example, for the demonstrated $LiNbO_3$ microring cavities, the strong second-order optical nonlinearity of $LiNbO_3$ predicts a single-photon nonlinear coupling strength of ~2.4 MHz (see Supplementary Information), which is attractive for optical quantum state engineering[33]. The PICs based on low-loss magneto-optic materials (e.g., YIG) will enable integrated nonreciprocal devices and also allow for the study of strong photon–magnon interaction[34]. Integrating laser crystals (e.g., BBO and YAG) with the BIC-based high-$Q$ low-modal-volume microcavities will create high-performance lasing or wavelength-conversion devices. Especially, the diamond containing the spin systems holds great potential for quantum memory and sensing at room temperature[35], and our PIC architecture can enhance the optical readout and control of the spins. By avoiding etching the high-refractive-index single-crystal materials, the BIC-based PICs have opened a new avenue in integrated photonics and also laid the foundation for exploring new optical functional materials on a chip which shall bring many unprecedented applications in classical and quantum photonics.

**Methods**

Methods including fabrication, characterization, and any associated accession codes are available in the Supplementary Information.

**References**

1. Longhi, S. Quantum-optical analogies using photonic structures. *Laser Photon. Rev.* **3**, 243–261 (2009).
2. Dragoman, D. & Dragoman, M. *Quantum-Classical Analogies*. (2013).
3. von Neumann, J. & Wigner, E. On some peculiar discrete eigenvalues. *Phys. Z.* **30**, 465–467 (1929).
4. Hsu, C. W., Zhen, B., Stone, A. D., Joannopoulos, J. D. & Soljačić, M. Bound states in the continuum. *Nat. Rev. Mater.* **1**, 16048 (2016).
5. Lyapina, A. A., Maksimov, D. N., Pilipchuk, A. S. & Sadreev, A. F. Bound states in the continuum in open acoustic resonators. *J. Fluid Mech.* **780**, 370–387 (2015).
6. Linton, C. M. & McIver, P. Embedded trapped modes in water waves and acoustics. *Wave Motion* **45**, 16–29 (2007).
7. Chen, Y. *et al.* Mechanical bound state in the continuum for optomechanical microresonators. *New J. Phys.* **18**, 063031 (2016).

8. Hein, S., Koch, W. & Nannen, L. Trapped modes and Fano resonances in two-dimensional acoustical duct–cavity systems. *J. Fluid Mech.* **692**, 257–287 (2012).

9. Xiao, Y.-X., Ma, G., Zhang, Z.-Q. & Chan, C. T. Topological subspace-induced bound state in the continuum. *Phys. Rev. Lett.* **118**, 166803 (2017).

10. Albo, A., Fekete, D. & Bahir, G. Electronic bound states in the continuum above (Ga,In)(As,N)/(Al,Ga)As quantum wells. *Phys. Rev. B* **85**, 115307 (2012).

11. Álvarez, C., Domínguez-Adame, F., Orellana, P. A. & Díaz, E. Impact of electron–vibron interaction on the bound states in the continuum. *Phys. Lett. A* **379**, 1062–1066 (2015).

12. Yan, J.-X. & Fu, H.-H. Bound states in the continuum and Fano antiresonance in electronic transport through a four-quantum-dot system. *Phys. B* **410**, 197–200 (2013).

13. Gong, W., Han, Y. & Wei, G. Antiresonance and bound states in the continuum in electron transport through parallel-coupled quantum-dot structures. *J. Phys. Condens. Matter* **21**, 175801 (2009).

14. Ladrón de Guevara, M. L. & Orellana, P. A. Electronic transport through a parallel-coupled triple quantum dot molecule: Fano resonances and bound states in the continuum. *Phys. Rev. B* **73**, 205303 (2006).

15. Capasso, F. *et al.* Observation of an electronic bound state above a potential well. *Nature* **358**, 565–567 (1992).

16. Gomis-Bresco, J., Artigas, D. & Torner, L. Anisotropy-induced photonic bound states in the continuum. *Nat. Photonics* **11**, 232–236 (2017).

17. Kodigala, A. *et al.* Lasing action from photonic bound states in continuum. *Nature* **541**, 196–199 (2017).

18. Hsu, C. W. *et al.* Observation of trapped light within the radiation continuum. *Nature* **499**, 188–191 (2013).

19. Zou, C.-L. *et al.* Guiding light through optical bound states in the continuum for ultrahigh-*Q* microresonators. *Laser Photon. Rev.* **9**, 114–119 (2015).

20. Marinica, D. C., Borisov, A. G. & Shabanov, S. V. Bound states in the continuum in photonics. *Phys. Rev. Lett.* **100**, 183902 (2008).

21. Plotnik, Y. *et al.* Experimental observation of optical bound states in the continuum. *Phys. Rev. Lett.* **107**, 183901 (2011).

22. Bulgakov, E. N. & Maksimov, D. N. Light guiding above the light line in arrays of dielectric nanospheres. *Opt. Lett.* **41**, 3888–3891 (2016).

23. Longhi, S. Optical analog of population trapping in the continuum: classical and quantum interference effects. *Phys. Rev. A* **79**, 023811 (2009).

24. Bulgakov, E. N. & Maksimov, D. N. Topological bound states in the continuum in arrays of dielectric spheres. *Phys. Rev. Lett.* **118**, 267401 (2017).

25. Zhen, B., Hsu, C. W., Lu, L., Stone, A. D. & Soljačić, M. Topological nature of optical bound states in the continuum. *Phys. Rev. Lett.* **113**, 257401 (2014).

26. Weimann, S. *et al.* Compact surface Fano states embedded in the continuum of waveguide arrays. *Phys. Rev. Lett.* **111**, 240403 (2013).

27. Monticone, F. & Alù, A. Embedded photonic eigenvalues in 3D nanostructures. *Phys. Rev. Lett.* **112**, 213903 (2014).

28. Rybin, M. V. *et al.* High-*Q* supercavity modes in subwavelength dielectric resonators. *Phys. Rev. Lett.* **119**, 243901 (2017).

29. Doeleman, H. M., Monticone, F., den Hollander, W., Alù, A. & Koenderink, A. F. Experimental observation of a polarization vortex at an optical bound state in the continuum. *Nat. Photonics* **12**, 397–401 (2018).

30. Silveirinha, M. G. Trapping light in open plasmonic nanostructures. *Phys. Rev. A* **89**, 023813 (2014).

31. Yu, Z., Cui, H. & Sun, X. Genetic-algorithm-optimized wideband on-chip polarization rotator with an ultrasmall footprint. *Opt. Lett.* **42**, 3093–3096 (2017).

32. Javerzac-Galy, C. *et al.* On-chip microwave-to-optical quantum coherent converter based on a superconducting resonator coupled to an electro-optic microresonator. *Phys. Rev. A* **94**, 053815 (2016).

33. Flayac, H. & Savona, V. Unconventional photon blockade. *Phys. Rev. A* **96**, 053810 (2017).

34. Zhang, X., Zhu, N., Zou, C.-L. & Tang, H. X. Optomagnonic whispering gallery microresonators. *Phys. Rev. Lett.* **117**, 123605 (2016).

35. Lončar, M. & Faraon, A. Quantum photonic networks in diamond. *MRS Bull.* **38**, 144–148 (2013).

**Acknowledgements**

This work was supported by the Early Career Scheme (24208915) and the General Research Fund (14208717, 14206318) sponsored by the Research Grants Council of Hong Kong, and by the NSFC/RGC Joint Research Scheme (N_CUHK415/15) sponsored by the Research Grants Council of Hong Kong and the National Natural Science Foundation of China. C.-L.Z. was supported by Anhui Initiative in Quantum Information Technologies (AHY130000).

**Authors contributions**

Z.Y. performed the theoretical modeling, numerical simulation, device design, fabrication, and characterization under the supervision of X.S.; X.X. assisted with device fabrication; C.-L.Z. provided theoretical guidance; Z.Y. and X.S. wrote the manuscript with the assistance of all the coauthors.

**Competing interests**

The authors declare no competing interests.

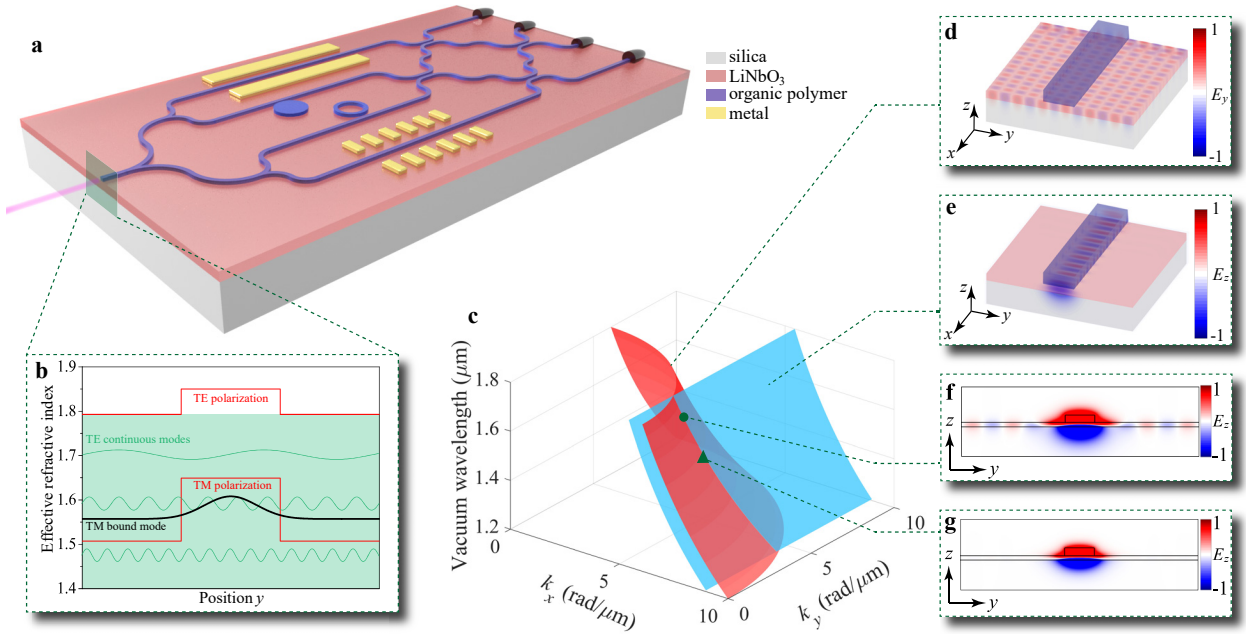

**Fig. 1 | BIC-based photonic integrated circuit and its modal properties. a,** Conceptual illustration of the proposed photonic circuit in the continuum. The bottom layer (gray) is silica substrate, the center layer (dark pink) is the single-crystal dielectric material (LiNbO$_3$), and the top layer consists of patterned low-index organic polymer (purple) and metal (yellow). The metal is used as electrodes for modulating guided photons at high speeds. **b,** Effective refractive index distributions for light propagating in the routing waveguides in **a** where the waveguide width $w$ = 1.8 μm. **c,** Dispersion diagram of the TM bound mode (blue surface) and the TE continuous modes (red surface). The intersecting line of the two surfaces represents phase matching of the TM bound mode and the TE continuous modes, where the former can be coupled with the latter and dissipate energy into the continuum. **d&e,** Electric field distributions of a TE continuous mode (**d**) and the TM bound mode (**e**) corresponding to the red and blue surfaces in **c**, respectively. **f&g,** Log-scale plots of the cross-sectional $E_z$ profiles of a general non-BIC mode (**f**) and a special BIC mode (**g**), which correspond to the green dot and green triangle on the intersecting line of the two surfaces in **c**.

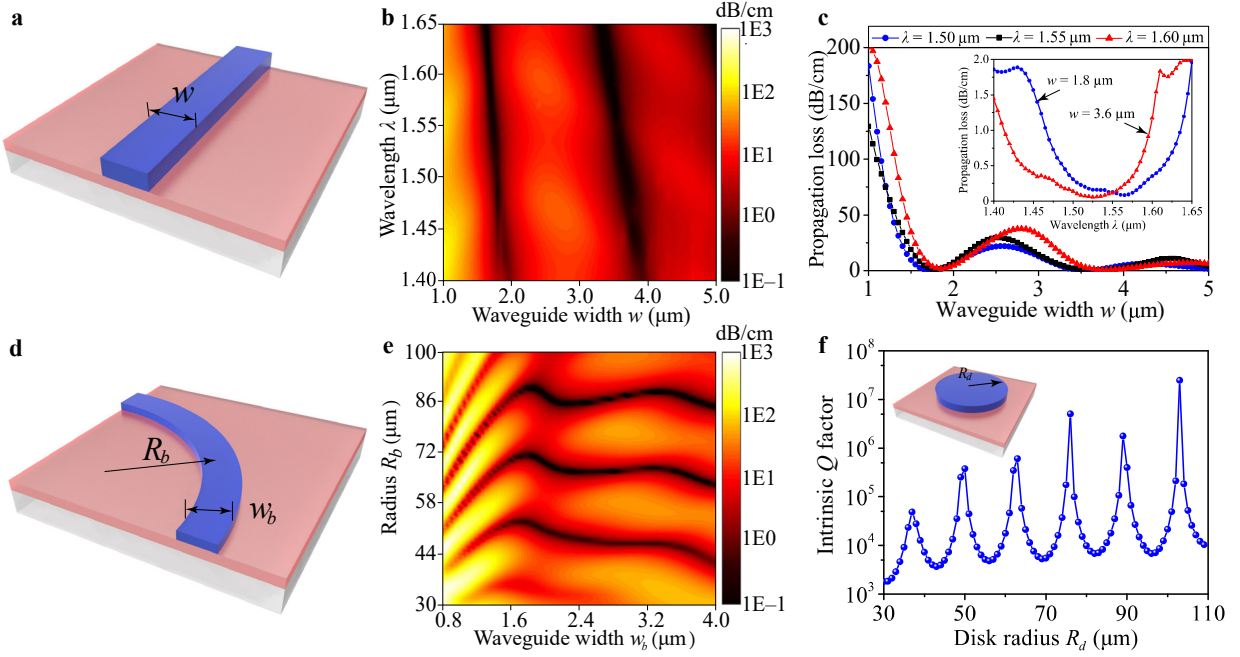

**Fig. 2 | Theoretical and numerical results of BICs. a,** Illustration of a straight waveguide in the continuum, where $w$ denotes the width of the waveguide. **b,** Propagation loss of the TM bound mode in the straight waveguide as a function of the width $w$ and wavelength $\lambda$. The propagation loss approaches zero (black regions) for certain combinations of parameters where the BIC condition is satisfied. **c,** Propagation loss as a function of the waveguide width $w$ for light at the wavelength of 1.50, 1.55, and 1.60 μm. The inset plots propagation loss as a function of the wavelength $\lambda$ with $w$ = 1.8 and 3.6 μm. Due to the low structural dispersion, the propagation loss can maintain below 2 dB/cm over a large bandwidth from 1.40 to 1.65 μm. **d,** Illustration of a bent waveguide, where $w_b$ and $R_b$ denote the width and radius of the bent waveguide, respectively. **e,** Bending loss as a function of the radius $R_b$ and width $w_b$. Similar to the straight waveguide, the bending loss approaches zero for certain combinations of parameters where the BIC condition is satisfied. **f,** Intrinsic $Q$ factor of microdisk cavities as a function of the disk radius $R_d$. The $Q$ factor approaches maxima that are limited only by the radiation loss for specific $R_d$ values where the BIC condition is satisfied.

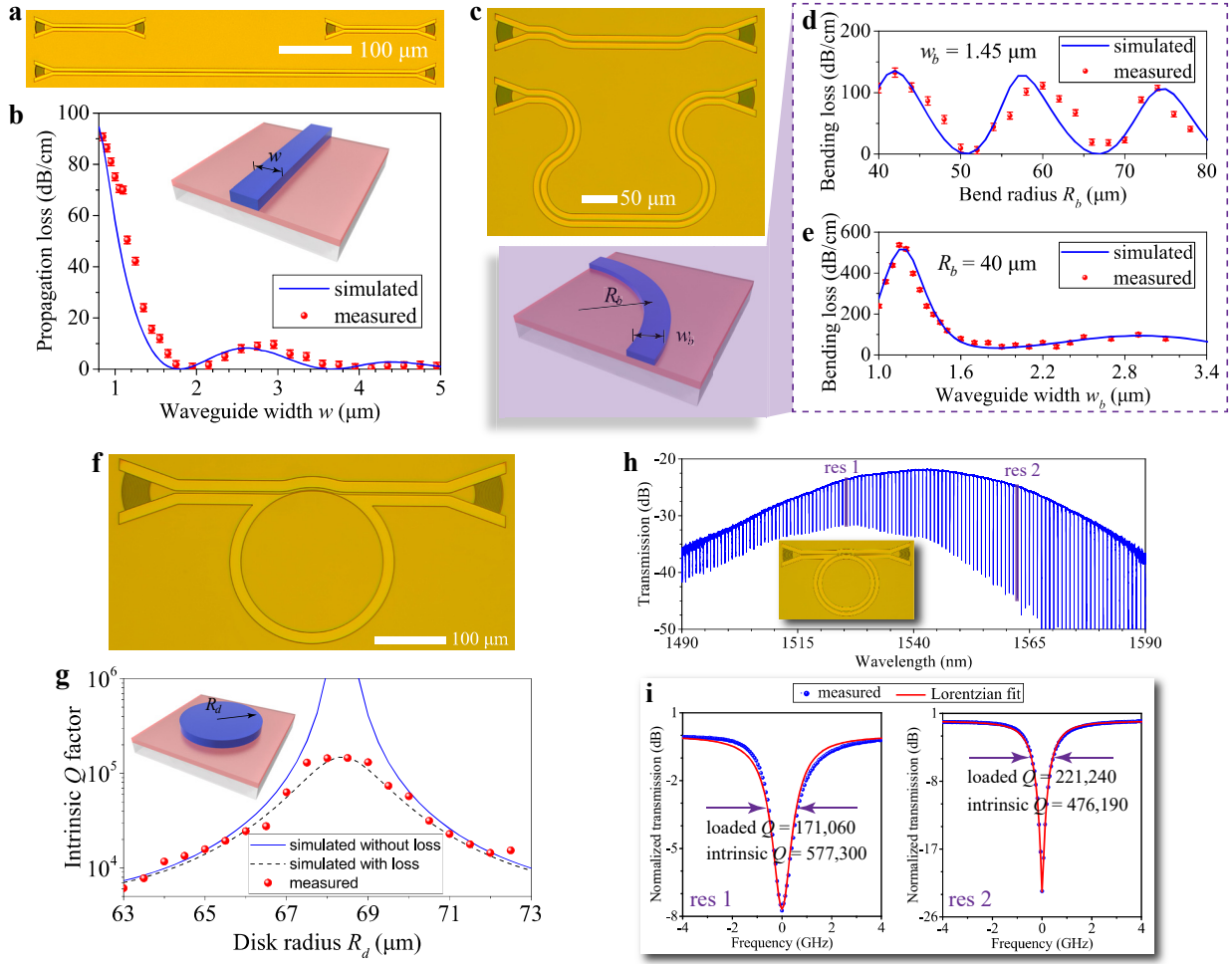

**Fig. 3 | Experimental demonstration of BICs. a,** Straight waveguides with in/output grating couplers. Devices with different waveguide lengths and identical in/output grating couplers were fabricated on the same chip to extract the propagation loss of the waveguides. **b,** Simulated and measured propagation loss of straight waveguides as a function of the width. **c,** Devices for measuring the bending loss in bent waveguides. The upper and lower devices have identical in/output grating couplers and the same length of straight waveguides, but different lengths of the bending sections. **d&e,** Simulated and measured bending loss as a function of the bend radius $R_b$ and width $w_b$. **f,** Microdisk cavity, evanescently coupled with a bus waveguide. **g,** Intrinsic $Q$ factors of the cavity resonances as a function of the disk radius (blue solid line: simulated curve

without consideration of loss; black dashed line: simulated curve with consideration of loss from material absorption and fabrication imperfection; red dots: measured data). **h,** Measured transmission spectrum of a microring cavity whose geometric parameters satisfy the BIC condition for light at 1,550 nm. Resonances of very high $Q$ factors with high extinction are achieved over a large wavelength range. **i,** Zoomed-in spectra of two typical resonances. The loaded and intrinsic $Q$ factors are $2.2 \times 10^5$ and $5.8 \times 10^5$, respectively, based on the Lorentzian fitting.

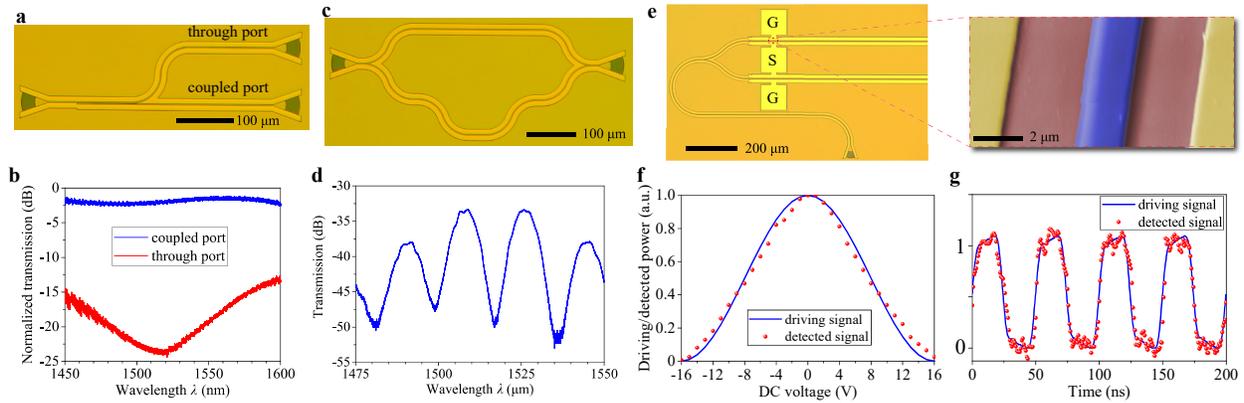

**Fig. 4 | Photonic devices based on BICs. a,** Optical microscope image of a directional coupler. **b,** Normalized transmission spectra of the coupled and through ports of the directional coupler, which show high coupling efficiency and high extinction ratio in the range of 1450 to 1600 nm. **c&d,** Optical microscope image and the measured transmission spectrum of a Mach–Zehnder interferometer (MZI). **e,** Optical microscope image of a MZI electro-optic modulator. The inset is a false-color SEM image zoomed in at the waveguide and the nearby electrodes. **f,** Measured optical output power of the MZI electro-optic modulator for light at the wavelength of 1,550 nm as a function of the applied DC voltage, indicating the $V_\pi$ of the modulator to be ~16 V. **g,** Temporal response of the MZI electro-optic modulator.